\documentclass{article}
\usepackage{amsfonts}
\usepackage{amssymb}
\usepackage{amsmath}
\usepackage{fancyhdr}
\usepackage[symbol]{footmisc}

\newtheorem{theorem}{Theorem}

\newtheorem{proposition}[theorem]{Proposition}

\newenvironment{proof}[1][Proof]{\noindent\textbf{#1.} }{\ \rule{0.5em}{0.5em}}
\input{tcilatex}
\begin{document}

\begin{center}
{\large \textbf{Jets, Lifts and Dynamics}}\\[1cm]
O\u{g}ul Esen\footnote{%
Corresponding author: phone:+902165781888, fax: +902165780672} and Hasan G%
\"{u}mral

Department of Mathematics, Yeditepe University

34755 Ata\c{s}ehir, Istanbul, Turkey

oesen@yeditepe.edu.tr \ \ \ \ \ hgumral@yeditepe.edu.tr \\[0.5cm]
\end{center}

\begin{abstract}
We show that complete cotangent lifts of vector fields, their decomposition
into vertical representative and holonomic part provide a geometrical
framework underlying Eulerian equations of continuum mechanics. We discuss
Euler equations for ideal incompressible fluid and Vlasov equations of
plasma dynamics in connection with the lifts of divergence-free and
Hamiltonian vector fields, respectively. As a further application, we obtain
kinetic equations of particles moving with the flow of contact vector fields
both from Lie-Poisson reductions and with the techniques of present
framework.

\end{abstract}%

\section{Jets}

Let $\left( \mathcal{E},\pi ,\mathcal{M}\right) $ be a smooth bundle with
coordinates $\left( x^{a};1\leq a\leq \dim \left( \mathcal{M}\right)
=m\right) $ on the base manifold $\mathcal{M}$ and $\left( x^{a},u^{\lambda
};1\leq \lambda \leq rank\left( \pi \right) =k\right) $ on the total
manifold $\mathcal{E}$. The vertical bundle associated with $\pi $ is 
\begin{equation}
V\pi =kerT\pi =\left\{ \xi \in T\mathcal{E}:T\pi \left( \xi \right)
=0\right\}   \label{vertical}
\end{equation}%
and this is a vector subbundle of the tangent bundle $T\mathcal{E}$. Here $%
T\pi $ denotes the tangent mapping of the projection $\pi $. Two sections $%
\phi ,\psi \in \mathfrak{S}\left( \pi \right) $ of the bundle $\pi $ at a
point $\mathbf{x}\in \mathcal{M}$ are called equivalent if their tangent
mappings are equal at that point, that is, $T_{x}\phi =T_{x}\psi .$ Given a
point $\mathbf{x}$, an equivalence class containing a section $\phi $ is
denoted by $j_{x}^{1}\phi .$ The first order jet manifold 
\begin{equation}
J^{1}\pi =\left\{ j_{x}^{1}\phi :\mathbf{x}\in \mathcal{M}\text{ and }\phi
\in \mathfrak{S}\left( \pi \right) \right\} 
\end{equation}%
associated with $\left( \mathcal{E},\pi ,\mathcal{M}\right) $ is the set of
equivalence classes at every point $\mathbf{x}\in \mathcal{M}$ with induced
coordinates 
\begin{equation}
\left( x^{a},u^{\lambda },u_{a}^{\lambda }\right) :J^{1}\pi \rightarrow 
\mathbb{R}
^{m+k+mk}:j_{x}^{1}\phi \rightarrow \left( x^{a},u^{\lambda }\left( \phi
\left( \mathbf{x}\right) \right) ,\left. \frac{\partial \phi ^{\lambda }}{%
\partial x^{a}}\right\vert _{x}\right) .
\end{equation}%
We have fibrations $\pi _{0}:J^{1}\pi \rightarrow \mathcal{E}:j_{x}^{1}\phi
\rightarrow \phi \left( \mathbf{x}\right) $ and\ $\pi _{1}:J^{1}\pi
\rightarrow \mathcal{M}:j_{x}^{1}\phi \rightarrow \mathbf{x}$ of $J^{1}\pi $
on $\mathcal{E}$ and $\mathcal{M}$, respectively \cite{kms93},\cite{sa}.

Given a differentiable map $\rho :\mathcal{N}\rightarrow \mathcal{M}$ from a
manifold $\mathcal{N}$ to the base manifold $\mathcal{M}$, the pull-back
bundle of $\pi $\ by $\rho $ is the triple $\left( \rho ^{\ast }\mathcal{E}%
,\rho ^{\ast }\pi ,\mathcal{N}\right) $ where 
\begin{equation}
\rho ^{\ast }\mathcal{E}=\mathcal{N}\times _{\mathcal{M}}\mathcal{E=}\left\{
\left( \mathbf{n,e}\right) \in \mathcal{N}\times \mathcal{E}:\pi \left( 
\mathbf{e}\right) =\rho \left( \mathbf{n}\right) \right\}
\end{equation}%
is the Whitney product and, $\rho ^{\ast }\pi =pr_{1}$ is the projection to
the first factor \cite{kms93}. Consider the pull back bundle 
\begin{equation*}
\left( \pi _{0}^{\ast }\left( T\mathcal{E}\right) =J^{1}\pi \times _{%
\mathcal{E}}T\mathcal{E},\pi _{0}^{\ast }\tau _{\mathcal{E}}=pr_{1},J^{1}\pi
\right)
\end{equation*}%
of $\left( T\mathcal{E},\tau _{\mathcal{E}},\mathcal{E}\right) $ by the
projection $\pi _{0}:J^{1}\pi \rightarrow \mathcal{E}$, where $\tau _{%
\mathcal{\mathcal{E}}}$ is the tangent bundle projection. A section of $\pi
_{0}^{\ast }\tau _{\mathcal{E}}$ is called a generalized vector field of
order one \cite{sa},\cite{Tul80}. One may regard a section of $\pi
_{0}^{\ast }\tau _{\mathcal{E}}$ as a map from $J^{1}\pi $ to $T\mathcal{E}.$
We require that generalized vector fields are projectable \cite{KoSc}.

In coordinates, a generalized vector field is 
\begin{equation}
\xi \left( j_{x}^{1}\phi \right) =\xi ^{a}\left( \mathbf{x}\right) \left. 
\frac{\partial }{\partial x^{a}}\right\vert _{x}+\xi ^{\lambda }\left(
j_{x}^{1}\phi \right) \left. \frac{\partial }{\partial u^{\lambda }}%
\right\vert _{\phi \left( x\right) }  \label{genvec}
\end{equation}%
and its first order prolongation $pr^{1}\xi $ is 
\begin{equation}
pr^{1}\xi =\xi +\Phi _{a}^{\lambda }\frac{\partial }{\partial u_{a}^{\lambda
}},\text{ \ \ }\Phi _{a}^{\lambda }=D_{x^{a}}\left( \xi ^{\lambda }-\xi
^{b}u_{b}^{\lambda }\right) +\xi ^{b}u_{ba}^{\lambda }
\end{equation}%
where $D_{x^{a}}$ is the total derivative operator with respect to $x^{a}$
and, $u_{ba}^{\lambda }\left( j_{x}\phi \right) =\partial ^{2}\phi ^{\lambda
}/\partial x^{a}\partial x^{b}$ is an element of the second order jet
bundle. Lie bracket of two first order generalized vector fields $\xi $ and $%
\eta $ is the unique first order generalized vector field 
\begin{equation}
\left[ \xi ,\eta \right] _{pro}=\left( pr^{1}\xi \left( \eta ^{a}\right)
-pr^{1}\eta \left( \xi ^{a}\right) \right) \frac{\partial }{\partial x^{a}}%
+\left( pr^{1}\xi \left( \eta ^{\lambda }\right) -pr^{1}\eta \left( \xi
^{\lambda }\right) \right) \frac{\partial }{\partial u^{\lambda }}.
\label{Liepro}
\end{equation}%
If $\xi $ and $\eta $ are two vector fields on $\mathcal{E}$, then $\left[ 
\text{ },\text{ }\right] _{pro}$ reduces to the Jacobi-Lie bracket of vector
fields \cite{olv86}.

\section{Lifts}

Consider a vector field $X\in \mathfrak{X}\left( \mathcal{M}\right) $ on $%
\mathcal{M}$, and let $\phi $ be a section of $\pi $. The holonomic lift of $%
X\left( \mathbf{x}\right) \in T_{x}\mathcal{M}$ by $\phi $ is 
\begin{equation}
\left( j_{x}^{1}\phi ,T\phi \left( X\left( \mathbf{x}\right) \right) \right)
\in \pi _{0}^{\ast }\left( T\mathcal{E}\right) =J^{1}\pi \times _{\mathcal{E}%
}T\mathcal{E}.
\end{equation}%
In coordinates, if $X=X^{a}\left( \mathbf{x}\right) \partial /\partial x^{a}$%
, then 
\begin{equation}
X^{hol}=X^{a}\frac{\partial }{\partial x^{a}}+X^{a}\frac{\partial \phi
^{\lambda }}{\partial x^{a}}\frac{\partial }{\partial u^{\lambda }}=X^{a}%
\frac{\partial }{\partial x^{a}}+X^{a}u_{a}^{\lambda }\left( j_{x}^{1}\phi
\right) \frac{\partial }{\partial u^{\lambda }}.
\end{equation}%
Define the holonomic part of a projectable vector field $\xi \in \mathfrak{X}%
\left( \mathcal{E}\right) $ as the holonomic lift of its push forward by $%
\pi $, that is 
\begin{equation}
H\xi =\left( \pi _{\ast }\xi \right) ^{hol}.  \label{holpart}
\end{equation}%
$H\xi $ is a generalized vector field of order one. Define a connection $%
(1;1)$ tensor 
\begin{equation}
\Gamma _{\mathbf{J}}\text{$=dx^{a}\otimes \left( \dfrac{\partial }{\partial
x^{a}}+u_{a}^{\lambda }\dfrac{\partial }{\partial u^{\lambda }}\right) .$}
\label{holcon}
\end{equation}%
satisfying $H\xi =\Gamma _{\mathbf{J}}\xi $. Then, the vertical (or
evolutionary) representative 
\begin{equation}
V\xi =\xi -\Gamma _{\mathbf{J}}\left( \xi \right) =\left( \xi ^{\alpha }-\xi
^{a}u_{a}^{\lambda }\right) \frac{\partial }{\partial u^{\lambda }}
\end{equation}%
of $\xi $ is vertical valued generalized vector field of order one \cite%
{olv86},\cite{sa},\cite{Tul80}.

\begin{proposition}
Holonomic lift is a Lie algebra isomorphism from the space of projectable
vector fields in $\mathfrak{X}(\mathcal{E})$ into $J^{1}\pi \times _{%
\mathcal{E}}T\mathcal{E}$.

\begin{proof}
We consider two projectable vector fields $\xi $ and $\eta $ on $\mathcal{E}$%
. A straight forward calculation gives 
\begin{equation}
\left[ \Gamma _{\mathbf{J}}\left( \xi \right) ,\Gamma _{\mathbf{J}}\left(
\eta \right) \right] _{pro}=\left[ \xi ^{hol},\eta ^{hol}\right] _{pro}=%
\left[ \xi ,\eta \right] ^{hol}=\Gamma _{\mathbf{J}}\left[ \xi ,\eta \right]
\end{equation}%
where $\left[ \text{ },\text{ }\right] _{pro}$ is the Lie bracket for
generalized vector fields in Eq.(\ref{Liepro}).
\end{proof}
\end{proposition}

On the other hand, the generalized bracket of vertical representatives
satisfies 
\begin{equation}
\left[ V\xi ,V\eta \right] _{pro}=V\left[ \xi ,\eta \right] _{pro}+\mathfrak{%
B}\left( \xi ,\eta \right) ,
\end{equation}%
where $\mathfrak{B}$ is a vertical-vector valued two-form 
\begin{equation}
\mathfrak{B}\left( \xi ,\eta \right) =\left[ \eta ^{hol},V\xi \right] _{pro}-%
\left[ \xi ^{hol},V\eta \right] _{pro}.  \label{Bi}
\end{equation}

There is, however, a class of vector fields, defined again by lifts, for
which the vertical representative becomes a Lie algebra isomorphism. Let $%
\varphi _{t}:\mathcal{M}\rightarrow \mathcal{M}$ be the flow of $X$ on $%
\mathcal{M}$. Cotangent lift of $\varphi _{t}$ is a one-parameter group of
diffeomorphism $\varphi _{t}^{c\ast }$ on $T^{\ast }\mathcal{M}$ satisfying%
\begin{equation}
\pi _{\mathcal{M}}\circ \varphi _{t}^{c\ast }=\varphi _{t}\circ \pi _{%
\mathcal{M}}  \label{cotanlift}
\end{equation}%
where $\pi _{\mathcal{M}}$ is the natural projection of $T^{\ast }\mathcal{M}
$ to $\mathcal{M}$. The cotangent lift of the inverse flow $T^{\ast }\varphi
_{-t}$ satisfies the argument in Eq.(\ref{cotanlift}). Infinitesimal
generator $X^{c\ast }:T^{\ast }\mathcal{M}\rightarrow TT^{\ast }\mathcal{M}$
of the flow $\varphi _{t}^{c\ast }$ is called complete cotangent lift of $X.$
$X^{c\ast }$ is a Hamiltonian vector field on the canonical symplectic
manifold $\left( T^{\ast }\mathcal{M},\Omega _{T^{\ast }\mathcal{M}%
}=-d\theta _{T^{\ast }\mathcal{M}}\right) $ for the Hamiltonian function $%
P\left( X\right) =i_{X^{c\ast }}\theta _{T^{\ast }\mathcal{M}}$ \cite{mr94}.
The infinitesimal version 
\begin{equation*}
T\pi _{\mathcal{M}}\circ X^{c\ast }=X\circ \pi _{\mathcal{M}}.
\end{equation*}%
of Eq.(\ref{cotanlift}) gives the relation between $X$ and $X^{c\ast }$ with 
$T\pi _{\mathcal{M}}$ being the tangent mapping of $\pi _{\mathcal{M}}$. The
complete cotangent lift mapping $^{c\ast }:\mathfrak{X}\left( \mathcal{M}%
\right) \rightarrow \mathfrak{X}\left( T^{\ast }\mathcal{M}\right) $ taking $%
X$ to $X^{c\ast }$ is a Lie algebra isomorphism into \cite{mr94},\cite{yano} 
\begin{equation}
\left[ X^{c\ast },Y^{c\ast }\right] =\left[ X,Y\right] ^{c\ast },\text{ \ \ }%
\forall X,Y\in \mathfrak{X}\left( \mathcal{M}\right) .  \label{cotlift}
\end{equation}

In Darboux's coordinates $\left( x^{a},y_{b}\right) $ on $T^{\ast }\mathcal{M%
},$ the complete cotangent lift of $X=X^{a}\left( \mathbf{x}\right) \partial
/\partial x^{a}$ on $\mathcal{M}$ is 
\begin{equation}
X^{c\ast }=X_{\mathcal{P}\left( X\right) }=X^{a}\dfrac{\partial }{\partial
x^{a}}-y_{b}\dfrac{\partial X^{b}}{\partial x^{a}}\dfrac{\partial }{\partial
y_{a}}  \label{ccl}
\end{equation}%
with the Hamiltonian function being $\mathcal{P}\left( X\right) \left( 
\mathbf{x},\mathbf{y}\right) =y_{b}X^{b}\left( \mathbf{x}\right) $. We
decompose the complete cotangent lifts into vertical representative and
holonomic part 
\begin{equation}
VX^{c\ast }=-(y_{b}\dfrac{\partial X^{b}}{\partial x^{a}}+X^{b}\frac{%
\partial y_{a}}{\partial x^{b}}){\frac{\partial }{\partial y_{a}}}\text{ \ \
and \ \ }HX^{c\ast }=X^{a}\dfrac{\partial }{\partial x^{a}}+X^{a}\frac{%
\partial y_{b}}{\partial x^{a}}{\frac{\partial }{\partial y_{b}}.}
\end{equation}%
where the connection in Eq.(\ref{holcon}) has the particular form%
\begin{equation}
\Gamma =\text{$dx^{a}\otimes \left( \dfrac{\partial }{\partial x^{a}}+\frac{%
\partial y_{b}}{\partial x^{a}}\dfrac{\partial }{\partial y_{b}}\right) .$}
\end{equation}

\begin{proposition}
The mapping $V^{c\ast }:\mathfrak{X}\left( \mathcal{M}\right) \rightarrow 
\mathfrak{X}\left( T^{\ast }\mathcal{M}\right) :X\rightarrow VX^{c\ast }$ is
a Lie algebra isomorphism into.

\begin{proof}
The vector valued two form $\mathfrak{B}$ in Eq.(\ref{Bi}) vanishes for the
complete cotangent lifts, that is, $\mathfrak{B}\left( X^{c\ast },Y^{c\ast
}\right) =0$ for all $X,Y\in \mathfrak{X}\left( \mathcal{M}\right) ,$
therefore one has $V\left[ X^{c\ast },Y^{c\ast }\right] =\left[ VX^{c\ast
},VY^{c\ast }\right] _{pro}$ and the result 
\begin{equation}
V\left[ X,Y\right] ^{c\ast }=\left[ VX^{c\ast },VY^{c\ast }\right] _{pro}.
\end{equation}%
follows from Eq.(\ref{cotlift}).
\end{proof}
\end{proposition}

The last object we consider in this section is the vertical lift of one
forms. Take the cotangent lift $T^{\ast }\pi _{\mathcal{M}}:T^{\ast }%
\mathcal{M}\rightarrow T^{\ast }T^{\ast }\mathcal{M}$ of the projection $\pi
_{\mathcal{M}}:T^{\ast }\mathcal{M}\rightarrow \mathcal{M}$ and recall the
isomorphism $\Omega _{T^{\ast }\mathcal{M}}^{\sharp }:T^{\ast }T^{\ast }%
\mathcal{M}\rightarrow TT^{\ast }\mathcal{M}$ associated with the symplectic
two-form $\Omega _{T^{\ast }\mathcal{M}}$ on $T^{\ast }\mathcal{M}.$ Define
the Euler vector field 
\begin{equation}
\mathcal{X}_{E}:T^{\ast }\mathcal{M}\rightarrow TT^{\ast }\mathcal{M}:%
\mathbf{z}\rightarrow \Omega _{T^{\ast }\mathcal{M}}^{\sharp }\circ T^{\ast
}\pi _{\mathcal{M}}\left( \mathbf{z}\right)  \label{ver}
\end{equation}%
which is vertical, that is, $image\left( \mathcal{X}_{E}\right) \subset
ker\left( T\pi _{\mathcal{M}}\right) $. Indeed, 
\begin{eqnarray}
\left\langle \mathbf{z},T\pi _{\mathcal{M}}\circ \mathcal{X}_{E}\left( 
\mathbf{z}\right) \right\rangle &=&\left\langle T^{\ast }\pi _{\mathcal{M}%
}\left( \mathbf{z}\right) \mathbf{,}\Omega _{T^{\ast }\mathcal{M}}^{\sharp
}\circ T^{\ast }\pi _{\mathcal{M}}\left( \mathbf{z}\right) \right\rangle 
\notag \\
&=&\Omega _{T^{\ast }\mathcal{M}}\left( T^{\ast }\pi _{\mathcal{M}}\left( 
\mathbf{z}\right) ,T^{\ast }\pi _{\mathcal{M}}\left( \mathbf{z}\right)
\right) =0,
\end{eqnarray}%
$\forall \mathbf{z}\in T^{\ast }\mathcal{M}$, where we used the
skew-symmetry of $\Omega _{T^{\ast }\mathcal{M}}$. $\mathcal{X}_{E}$ is the
unique vector field satisfying the following equalities 
\begin{equation}
i_{\mathcal{X}_{E}}\Omega _{T^{\ast }\mathcal{M}}=\theta _{T^{\ast }\mathcal{%
M}},\text{ \ \ }\mathcal{L}_{\mathcal{X}_{E}}\Omega _{T^{\ast }\mathcal{M}%
}=-\Omega _{T^{\ast }\mathcal{M}}\text{, \ \ }\mathcal{L}_{\mathcal{X}%
_{E}}\theta _{T^{\ast }\mathcal{M}}=-\theta _{T^{\ast }\mathcal{M}},
\label{verform}
\end{equation}%
where $i_{\mathcal{X}_{E}}$ and $\mathcal{L}_{\mathcal{X}_{E}}$ are the
interior product and the Lie derivative operators \cite{LiMa}. Let $\alpha
\in \Lambda ^{1}\left( \mathcal{M}\right) $ be a one-form on $\mathcal{M}$.
The vertical lift 
\begin{equation}
\alpha ^{v}=\mathcal{X}_{E}\circ \alpha \circ \pi _{\mathcal{M}}:T^{\ast }%
\mathcal{M}\rightarrow TT^{\ast }\mathcal{M}
\end{equation}%
of the one-from $\alpha $ is a vertical vector field on $T^{\ast }\mathcal{M}
$. The Jacobi-Lie bracket of a complete cotangent lift and a vertical lift
is a vertical lift 
\begin{equation}
\left[ X^{c\ast },\alpha ^{v}\right] =\left( \mathcal{L}_{X}\alpha \right)
^{v}
\end{equation}%
for $X\in \mathfrak{X}\left( \mathcal{M}\right) $ and $\alpha \in \Lambda
^{1}\left( \mathcal{M}\right) $ \cite{yano}. In coordinates $\left(
x^{a},y_{b}\right) $ of $T^{\ast }\mathcal{M}$, the Euler vector field is $%
\mathcal{X}_{E}=-y_{a}\partial /\partial y_{a}$ and the vertical lift of the
one-form $\alpha =\alpha _{a}\left( \mathbf{x}\right) dx^{a}$ becomes $%
\alpha ^{v}=-\alpha _{a}\left( \mathbf{x}\right) \partial /\partial y_{a}.$

\section{Dynamics}

Assume that a continuum initially rests in $\mathcal{M}$, and the group $%
Diff\left( \mathcal{M}\right) $ of diffeomorphisms acts on left by
evaluation on $\mathcal{M}$ 
\begin{equation}
Diff\left( \mathcal{M}\right) \times \mathcal{M}\rightarrow \mathcal{M}%
:(\varphi ,\mathbf{x})\rightarrow \varphi \left( \mathbf{x}\right)
\label{ga}
\end{equation}%
to produce the motion of particles. The right action of $Diff\left( \mathcal{%
M}\right) $ commutes with the particle motion and constitutes an infinite
dimensional symmetry group of the kinematical description. This is the
particle relabelling symmetry \cite{arkh}. An element of the tangent space $%
T_{\varphi }Diff\left( \mathcal{M}\right) $ at $\varphi \in Diff\left( 
\mathcal{M}\right) $ is a map $V_{\varphi }:\mathcal{M}\rightarrow T\mathcal{%
M}$ called the material velocity field\textbf{\ }and satisfies $\tau _{%
\mathcal{M}}\circ V_{\varphi }=\varphi $. In particular, the tangent space $%
T_{id_{\mathcal{M}}}Diff\left( \mathcal{M}\right) $ at the identity $id_{%
\mathcal{M}}\in Diff\left( \mathcal{M}\right) $ is the space $\mathfrak{X}%
\left( \mathcal{M}\right) $ of smooth vector fields on $\mathcal{M}$. The
Lie algebra of $Diff(\mathcal{M})$ is $\mathfrak{X}\left( \mathcal{M}\right) 
$ with minus the Jacobi-Lie bracket of vector fields \cite{mr94}.

The dual space $\mathfrak{X}^{\ast }\left( \mathcal{M}\right) \simeq \Lambda
^{1}\left( \mathcal{M}\right) \otimes Den\left( \mathcal{M}\right) $ of the
Lie algebra is the space of one-form densities on $\mathcal{M}$. The pairing
between $\alpha \otimes d\mu \in \mathfrak{X}^{\ast }\left( \mathcal{M}%
\right) $ and $X\in \mathfrak{X}\left( \mathcal{M}\right) $ is given by%
\begin{equation}
\left\langle \alpha \otimes d\mu ,X\right\rangle =\int_{\mathcal{M}%
}\left\langle \alpha \left( \mathbf{x}\right) ,X\left( \mathbf{x}\right)
\right\rangle d\mu \left( \mathbf{x}\right) .  \label{pairing}
\end{equation}%
The pairing inside the integral is the natural pairing of finite dimensional
spaces $T_{x}\mathcal{M}$ and $T_{x}^{\ast }\mathcal{M}$. The coadjoint
action is 
\begin{eqnarray}
ad_{X}^{\ast } &:&\mathfrak{X}^{\ast }\left( \mathcal{M}\right) \rightarrow 
\mathfrak{X}^{\ast }\left( \mathcal{M}\right)  \notag  \label{coadjdiff} \\
&:&\alpha \otimes d\mu \rightarrow \mathcal{L}_{X}\left( \alpha \otimes d\mu
\right) =\left( \mathcal{L}_{X}\alpha +\left( div_{d\mu }X\right) \alpha
\right) \otimes d\mu
\end{eqnarray}%
$\forall X\in \mathfrak{X}\left( \mathcal{M}\right) $ and hence the
Lie-Poisson equations on $\mathfrak{X}^{\ast }\left( \mathcal{M}\right) $
are 
\begin{equation}
\dot{\alpha}=-\mathcal{L}_{X}\alpha -\left( div_{d\mu }X\right) \alpha ,
\label{LP}
\end{equation}%
where $div_{d\mu }X$ denotes the divergence of the vector field $X$ with
respect to the volume form $d\mu $.

In terms of vertical lifts, the dynamics in Eq.(\ref{LP}) is generated by
the vector field $\left( \mathcal{L}_{X}\alpha +\left( div_{d\mu }X\right)
\alpha \right) ^{v}$. For the divergence free vector fields, if $\alpha
=y_{a}dx^{a}$, then the Lie-Poisson equations are generated by 
\begin{equation}
(\mathcal{L}_{X}(y_{a}dx^{a}))^{v}=VX^{c\ast }\left( x^{a},y_{a}\right) .
\end{equation}

\subsection{Ideal incompressible fluid}

For an ideal incompressible fluid in a bounded compact region $\mathcal{%
Q\subset 
\mathbb{R}
}^{3}$ the configuration space is the group $Diff_{vol}\left( \mathcal{Q}%
\right) $ of volume preserving diffeomorphisms on $\mathcal{Q}$. The Lie
algebra $\mathfrak{X}_{div}\left( \mathcal{Q}\right) $ of $Diff_{vol}\left( 
\mathcal{Q}\right) $ is the algebra of divergence free vector fields
parallel to the boundary of $\mathcal{Q}$ and, the dual space $\mathfrak{X}%
_{div}^{\ast }\left( \mathcal{Q}\right) $ is the space 
\begin{equation}
\mathfrak{X}_{div}^{\ast }\left( \mathcal{Q}\right) =\{\left[ \Upsilon %
\right] \otimes d^{3}\mathbf{q}\in (\Lambda ^{1}(\mathcal{Q})/d\mathcal{F}(%
\mathcal{Q}))\otimes Den(\mathcal{Q})\},
\end{equation}%
of one-form modulo exact one-form densities on $\mathcal{Q}$. Here, $\left[
\Upsilon \right] =\left\{ \Upsilon +d\tilde{p}:\tilde{p}\text{ }\in \text{ }%
\mathcal{F}(\mathcal{Q})\right\} $ denotes the equivalence class containing $%
\Upsilon $ and the volume three form $d^{3}\mathbf{q}$ is the Euclidean
volume on $\mathcal{%
\mathbb{R}
}^{3}$ \cite{arkh},\cite{mw83}.

Let $\left( x^{a},\Upsilon _{b}\right) $ be induced coordinates and $%
X=X^{a}\partial /\partial x^{a}$ be a divergence free vector field. The
complete cotangent lift of $X$ is 
\begin{equation*}
X^{c\ast }=X^{a}\dfrac{\partial }{\partial x^{a}}-\Upsilon _{b}\left(
\partial X^{b}/\partial x^{a}\right) \dfrac{\partial }{\partial \Upsilon _{a}%
}
\end{equation*}%
and its vertical representative becomes 
\begin{equation}
VX^{c\ast }=\left( -\Upsilon _{b}\dfrac{\partial X^{b}}{\partial x^{a}}-X^{a}%
\dfrac{\partial \Upsilon _{b}}{\partial x^{a}}\right) \dfrac{\partial }{%
\partial \Upsilon _{a}}.
\end{equation}%
Equations of motion for the dynamics generated by $VX^{c\ast }$ are 
\begin{equation}
\frac{\partial \left[ \Upsilon \right] }{\partial t}=-\mathcal{L}_{X}\left[
\Upsilon \right] .  \label{preEuler}
\end{equation}%
For a generic element $\Upsilon +d\tilde{p}\in \left[ \Upsilon \right] ,$
Eq.(\ref{preEuler}) becomes Euler's equations for ideal fluid, that is $%
\partial \Upsilon /\partial t+\mathcal{L}_{X}\Upsilon =dp.$ If the dual
space $\mathfrak{X}_{div}^{\ast }\left( \mathcal{Q}\right) $ is identified
with exact two forms by $\left[ \Upsilon \right] \rightarrow d\Upsilon
=\omega \in \Lambda ^{2}(\mathcal{Q})$, then Eq.(\ref{preEuler}) becomes the
Euler's equation in vorticity form $\partial \omega /\partial t+\mathcal{L}%
_{X}\omega =0.$

\subsection{Collisionless plasma}

We take $\mathcal{M}$ to be cotangent bundle $T^{\ast }\mathcal{Q}$ of $%
\mathcal{Q}\subset 
\mathbb{R}
^{3}$ in which the plasma particles move. The configuration space of
collisionless nonrelativistic plasma is the group 
\begin{equation}
Diff_{can}\left( T^{\ast }\mathcal{Q}\right) =\left\{ \varphi \in T^{\ast }%
\mathcal{Q}:\varphi ^{\ast }\Omega _{T^{\ast }\mathcal{Q}}=\Omega _{T^{\ast }%
\mathcal{Q}}\right\} 
\end{equation}%
of all canonical diffeomorphisms where $\Omega _{T^{\ast }\mathcal{Q}}$ is
the canonical symplectic two form on $T^{\ast }\mathcal{Q}$ \cite{gpd1},\cite%
{mwrss83},\cite{mw82}. We assume that, the Lie algebra of $Diff_{can}\left(
T^{\ast }\mathcal{Q}\right) $ is the space of globally Hamiltonian vector
fields $\mathfrak{X}_{ham}\left( T^{\ast }\mathcal{Q}\right) $ with minus
the Jacobi-Lie bracket so that the equations 
\begin{equation}
\lbrack X_{h},X_{f}]_{JL}=-X_{\{h,f\}_{\Omega _{T^{\ast }\mathcal{Q}}}}
\label{Poissham}
\end{equation}%
describe a Lie algebra isomorphism 
\begin{equation}
h\rightarrow X_{h}:\left( \mathcal{F}\left( T^{\ast }\mathcal{Q}\right) ,\{%
\text{ },\text{ }\}_{\Omega _{T^{\ast }\mathcal{Q}}}\right) \rightarrow
\left( \mathfrak{X}_{ham}\left( T^{\ast }\mathcal{Q}\right) ,-[\text{ },%
\text{ }]_{JL}\right) ,  \label{iso1}
\end{equation}%
between $\mathfrak{X}_{ham}\left( T^{\ast }\mathcal{Q}\right) $ and the
space of smooth functions $\mathcal{F}\left( T^{\ast }\mathcal{Q}\right) $
modulo constants endowed with the (nondegenerate) canonical Poisson bracket $%
\{$ $,$ $\}_{\Omega _{T^{\ast }\mathcal{Q}}}$.

\begin{proposition}
The dual space of the Lie algebra $\mathfrak{X}_{ham}\left( T^{\ast }%
\mathcal{Q}\right) $ of Hamiltonian vector fields is 
\begin{equation}
\mathfrak{X}_{ham}^{\ast }\left( T^{\ast }\mathcal{Q}\right) =\{\Pi
_{id}\otimes d\mu \in \Lambda ^{1}(T^{\ast }\mathcal{Q})\otimes Den(T^{\ast }%
\mathcal{Q}):div_{\Omega _{T^{\ast }\mathcal{Q}}}\Pi _{id}^{\sharp }\neq 0\}.
\label{momdef}
\end{equation}
\end{proposition}

With this definition of the dual space the $L_{2}$-pairing of the Lie
algebra and its dual becomes nondegenerate provided we take the volume form
to be the symplectic one $d\mu =\Omega _{T^{\ast }\mathcal{Q}}^{3}$ in 
\begin{eqnarray}
\int_{T^{\ast }\mathcal{Q}}\left\langle X_{h}\left( \mathbf{z}\right) ,\Pi
_{id}\left( \mathbf{z}\right) \right\rangle d\mu \left( \mathbf{z}\right) 
&=&-\int_{T^{\ast }\mathcal{Q}}\left\langle dh,\Pi _{id}^{\sharp
}\right\rangle d\mu =-\int_{T^{\ast }\mathcal{Q}}i_{\Pi _{id}^{\sharp
}}\left( dh\right) d\mu   \notag \\
&=&-\int_{T^{\ast }\mathcal{Q}}dh\wedge i_{\Pi _{id}^{\sharp }}d\mu
=\int_{T^{\ast }\mathcal{Q}}hdi_{\Pi ^{\sharp }}d\mu   \notag \\
&=&\int_{T^{\ast }\mathcal{Q}}hdiv_{\Omega _{T^{\ast }\mathcal{Q}}}\Pi
_{id}^{\sharp }d\mu ,
\end{eqnarray}%
where we use the musical isomorphism $\Omega _{T^{\ast }\mathcal{Q}}^{\sharp
}:\Pi _{id}\rightarrow \Pi _{id}^{\sharp }$ induced from the symplectic
two-form $\Omega _{T^{\ast }\mathcal{Q}}$ and apply integration by parts 
\cite[internet supplement]{mr94}. The dual of the Lie algebra isomorphism in
Eq.(\ref{iso1}) is%
\begin{equation}
\Pi _{id}\left( \mathbf{z}\right) \rightarrow div_{\Omega _{T^{\ast }Q}}\Pi
_{id}^{\sharp }\left( \mathbf{z}\right)   \label{mom1}
\end{equation}%
and it is a momentum map. In Darboux's coordinates $\mathbf{z}=\left(
q^{i},p_{i}\right) $ on $T^{\ast }\mathcal{Q}$, we have $\Omega _{T^{\ast
}Q}=dq^{i}\wedge dp_{i}$ and\ we take $\Pi _{id}=\Pi _{i}\left( \mathbf{z}%
\right) dq^{i}+\Pi ^{i}\left( \mathbf{z}\right) dp_{i}$.$\ $Then, the
momentum map%
\begin{equation}
f\left( \mathbf{z}\right) =div_{\Omega _{T^{\ast }Q}}\Pi _{id}^{\sharp
}\left( \mathbf{z}\right) =\frac{\partial \Pi ^{i}\left( \mathbf{z}\right) }{%
\partial q^{i}}-\frac{\partial \Pi _{i}\left( \mathbf{z}\right) }{\partial
p_{i}}  \label{ede}
\end{equation}%
defines the plasma density function.

In the induced coordinates $\left( q^{i},p_{j};\Pi _{i},\Pi ^{j}\right) $ on 
$T^{\ast }T^{\ast }\mathcal{Q}$, consider the Hamiltonian function $h=\left(
1/2m\right) \delta ^{ij}p_{i}p_{j}+e\phi \left( \mathbf{q}\right) $ which is
the energy of a charged particle on $\mathcal{Q}$ \cite{mwrss83}. The
corresponding Hamiltonian vector field is 
\begin{equation}
X_{h}(\mathbf{z})=\frac{1}{m}\delta ^{ij}p_{i}\frac{\partial }{\partial q^{j}%
}-e\frac{\partial \phi }{\partial q^{i}}\frac{\partial }{\partial p_{i}}.
\end{equation}%
The complete cotangent lift of $X_{h}$ and its decomposition into vertical
representative and holonomic part are%
\begin{eqnarray}
X_{h}^{c\ast } &=&X_{h}-\delta ^{ij}{\frac{1}{m}}\Pi _{i}\frac{\partial }{%
\partial \Pi ^{j}}+e\Pi ^{j}\frac{\partial ^{2}\phi }{\partial q^{j}\partial
q^{i}}\frac{\partial }{\partial \Pi _{i}},  \notag \\
HX_{h}^{c\ast } &=&X_{h}+X_{h}\left( \Pi _{i}\right) \frac{\partial }{%
\partial \Pi _{i}}+X_{h}\left( \Pi ^{i}\right) \frac{\partial }{\partial \Pi
^{i}},  \notag \\
VX_{h}^{c\ast } &=&\left( e\Pi ^{j}{\frac{\partial ^{2}\phi }{\partial
q^{j}\partial q^{i}}}-X_{h}(\Pi _{i})\right) {\frac{\partial }{\partial \Pi
_{i}}-}({\frac{1}{m}}\Pi _{j}\delta ^{ji}+X_{h}(\Pi ^{i})){\frac{\partial }{%
\partial \Pi ^{i}},}
\end{eqnarray}%
where $X_{h}\left( \Pi _{i}\right) $ denotes the action of $X_{h}$ on $\Pi
_{i}$. Since, Hamiltonian vector fields are divergence free, the Lie-Poisson
equations 
\begin{eqnarray}
\dot{\Pi}_{i} &=&-X_{h}\left( \Pi _{i}\right) +e\frac{\partial ^{2}\phi }{%
\partial q^{i}\partial q^{j}}\Pi ^{j}  \notag \\
\dot{\Pi}^{i} &=&-X_{h}\left( \Pi ^{i}\right) -\frac{1}{m}\delta ^{ij}\Pi
_{j}  \label{mv}
\end{eqnarray}%
are generated solely by $VX_{h}^{c\ast }$. These are Vlasov equations in the
momentum variables \cite{gpd1}. For the density formulation, we make
back-substitution of the plasma density function $f\left( \mathbf{z}\right)
=div_{\Omega _{T^{\ast }Q}}\Pi _{id}^{\sharp }$ into Eqs. (\ref{mv}) and
obtain the Vlasov equation 
\begin{equation}
\frac{\partial f}{\partial t}+\frac{\delta ^{ij}p_{i}}{m}\frac{\partial f}{%
\partial q^{j}}-e\frac{\partial \phi }{\partial q^{i}}\frac{\partial f}{%
\partial p_{i}}=0.
\end{equation}

\subsection{Contact flows in $3D$}

Let $\mathcal{M}$ be a three dimensional manifold with a contact one form $%
\sigma \in \Lambda ^{1}\left( \mathcal{M}\right) $ satisfying $d\sigma
\wedge \sigma \neq 0.$ A contact form determines a contact structure which,
locally is the kernel of the contact form $\sigma $. A diffeomorphism on $%
\mathcal{M}$ is called a contact diffeomorphism if it preserves the contact
structure. We denote the group of contact diffeomorphisms by $%
Diff_{con}\left( \mathcal{M}\right) $. A vector field on a contact manifold $%
\left( \mathcal{M},\sigma \right) $ is called a contact vector field if it
generates a one-parameter group of contact diffeomorphisms \cite{arn89},\cite%
{ms98}.

In Darboux's coordinates $\left( x,y,z\right) $ on $\mathcal{M}$, we take
the contact form to be $\sigma =xdy+dz.$ For a real valued function $%
K=K\left( x,y,z\right) $ on $\mathcal{M}$, there corresponds a contact
vector field 
\begin{equation}
X_{K}=\left( \frac{\partial K}{\partial y}-x\frac{\partial K}{\partial z}%
\right) \frac{\partial }{\partial x}-\frac{\partial K}{\partial x}\frac{%
\partial }{\partial y}+\left( -K+x\frac{\partial K}{\partial x}\right) \frac{%
\partial }{\partial z},  \label{convec}
\end{equation}%
on $\mathcal{M}$ satisfying the identities%
\begin{equation}
i_{X_{K}}\sigma =-K\text{ \ \ and \ \ }i_{X_{K}}d\sigma =dK-\left(
i_{R_{\sigma }}dK\right) \sigma ,  \label{contact}
\end{equation}%
where $R_{\sigma }=\partial /\partial z$ is the Reeb vector field of $\sigma
.$ $R_{\sigma }$ is the unique vector field satisfying $i_{R_{\sigma
}}\sigma =1$ and $i_{R_{\sigma }}d\sigma =0.$ The divergence $div_{d\mu
}X_{K}$ of $X_{K}$ with respect to the volume form $d\mu =d\sigma \wedge
\sigma $ can be computed to be $div_{d\mu }X_{K}=-2R_{\sigma }K$.

Contact Poisson (or Lagrange) bracket of two smooth functions on $\mathcal{M}
$ is defined by%
\begin{equation}
\left\{ L,K\right\} _{c}=\frac{\partial L}{\partial x}\frac{\partial K}{%
\partial y}-\frac{\partial L}{\partial y}\frac{\partial K}{\partial x}+\frac{%
\partial K}{\partial z}\left( L-x\frac{\partial L}{\partial x}\right) -\frac{%
\partial L}{\partial z}\left( K-x\frac{\partial K}{\partial x}\right) ,
\end{equation}%
$\forall L,K\in \mathcal{F}\left( \mathcal{M}\right) .$ The identity $\left[
X_{K},X_{L}\right] _{JL}=-X_{\left\{ K,L\right\} _{c}}$ establishes an
isomorphism between Lie algebras $\left( \mathfrak{X}_{con}\left( \mathcal{M}%
\right) ,-\left[ \text{ },\text{ }\right] _{JL}\right) $ and $\left( 
\mathcal{F}\left( \mathcal{M}\right) ,\left\{ \text{ },\text{ }\right\}
_{c}\right) $. Following result gives a precise definition of the linear
algebraic dual of $\mathfrak{X}_{con}\left( \mathcal{M}\right) $.

\begin{proposition}
The dual space of the algebra $\mathfrak{X}_{con}\left( \mathcal{M}\right) $
of contact vector fields is 
\begin{equation}
\mathfrak{X}_{con}^{\ast }\left( \mathcal{M}\right) =\left\{ \alpha \otimes
d\mu \in \Lambda ^{1}\left( \mathcal{M}\right) \otimes Den\left( \mathcal{M}%
\right) :d\alpha \wedge \sigma -2\alpha \wedge d\sigma \neq 0\right\}
\label{condual}
\end{equation}%
where $\sigma $ is the contact form on $\mathcal{M}$ and $d\mu =d\sigma
\wedge \sigma $.

\begin{proof}
\begin{proof}
This follows from the requirement that the pairing between $\mathfrak{X}%
_{con}\left( \mathcal{M}\right) $ and $\mathfrak{X}_{con}^{\ast }\left( 
\mathcal{M}\right) $ be nondegenerate. We compute 
\begin{eqnarray}
\int_{\mathcal{M}}\left\langle \alpha ,X_{K}\right\rangle d\mu  &=&\int_{%
\mathcal{M}}\alpha \wedge i_{X_{K}}d\sigma \wedge \sigma +\int_{\mathcal{M}%
}\left( i_{X_{K}}\sigma \right) \alpha \wedge d\sigma   \notag \\
&=&\int_{\mathcal{M}}\alpha \wedge \left( dK-\left( i_{R_{\sigma }}dK\right)
\sigma \right) \wedge \sigma -\int_{\mathcal{M}}K\alpha \wedge d\sigma  
\notag \\
&=&\int_{\mathcal{M}}K\left( d\alpha \wedge \sigma -2\alpha \wedge d\sigma
\right) ,
\end{eqnarray}%
where we use the identities in Eq.(\ref{contact}) at the second step.
\end{proof}
\end{proof}
\end{proposition}

A geometric definition of density of contact particles can be achieved by
considering the Lie algebra isomorphism $\mathcal{F}\left( \mathcal{M}%
\right) \rightarrow \mathfrak{X}_{con}\left( \mathcal{M}\right)
:K\rightarrow X_{K}$ the dual of which is a momentum map 
\begin{equation}
\mathfrak{X}_{con}^{\ast }\left( \mathcal{M}\right) \rightarrow Den\left( 
\mathcal{M}\right) :\alpha \rightarrow d\alpha \wedge \sigma -2\alpha \wedge
d\sigma .  \label{condenmom}
\end{equation}%
and defines a real valued function $L$ on $\mathcal{M}$ 
\begin{equation}
Ld\sigma \wedge \sigma =d\alpha \wedge \sigma -2\alpha \wedge d\sigma .
\end{equation}%
In coordinates, let $\alpha =\alpha _{x}dx+\alpha _{y}dy+\alpha _{z}dz\in 
\mathfrak{X}_{con}^{\ast }\left( \mathcal{M}\right) $ and recall $\sigma
=xdy+dz$. Then, 
\begin{equation}
L(x,y,z)=-\frac{\partial \alpha _{x}}{\partial y}+\frac{\partial \alpha _{y}%
}{\partial x}-x\frac{\partial \alpha _{z}}{\partial x}+x\frac{\partial
\alpha _{x}}{\partial z}-2\alpha _{z}.  \label{el}
\end{equation}

The dual space $\mathfrak{X}_{con}^{\ast }\left( \mathcal{N}\right) $ admits
the Lie-Poisson bracket 
\begin{equation}
\left\{ \mathfrak{H},\mathfrak{K}\right\} \left( \alpha \right) =-\int_{%
\mathcal{M}}\left\langle \alpha ,\left[ \frac{\delta \mathfrak{H}}{\delta
\alpha },\frac{\delta \mathfrak{K}}{\delta \alpha }\right]
_{JL}\right\rangle d\mu =-\int_{\mathcal{M}}\left\langle \alpha ,\left[
X_{H},X_{K}\right] _{JL}\right\rangle d\mu ,  \label{conLP}
\end{equation}%
where $\mathfrak{H},\mathfrak{K}\in \mathcal{F}\left( \mathfrak{X}%
_{con}^{\ast }\left( \mathcal{M}\right) \right) $ and $\delta \mathfrak{H/}%
\delta \alpha =X_{H},$ $\delta \mathfrak{K/}\delta \alpha =X_{K}\in 
\mathfrak{X}_{con}\left( \mathcal{M}\right) $. The Hamiltonian operator $%
J_{LP}\left( \alpha \right) $ associated to the Lie-Poisson bracket in Eq.(%
\ref{conLP}) is defined by 
\begin{equation}
\left\{ \mathfrak{H},\mathfrak{K}\right\} \left( \alpha \right) =-\int_{%
\mathcal{M}}\left\langle X_{H},J_{LP}\left( \alpha \right)
X_{K}\right\rangle d\mu 
\end{equation}%
and a direct computation gives

\begin{proposition}
The Hamiltonian differential operator associated to the Lie-Poisson bracket
in Eq.(\ref{conLP}) is 
\begin{equation}
J_{LP}\left( \alpha \right) =-%
\begin{pmatrix}
\alpha _{x}\dfrac{\partial }{\partial x}+\dfrac{\partial }{\partial x}\cdot
\alpha _{x} & \alpha _{y}\dfrac{\partial }{\partial x}+\dfrac{\partial }{%
\partial y}\cdot \alpha _{x} & \alpha _{z}\dfrac{\partial }{\partial x}+%
\dfrac{\partial }{\partial z}\cdot \alpha _{x} \\ 
\alpha _{x}\dfrac{\partial }{\partial y}+\dfrac{\partial }{\partial x}\cdot
\alpha _{y} & \alpha _{y}\dfrac{\partial }{\partial y}+\dfrac{\partial }{%
\partial y}\cdot \alpha _{y} & \alpha _{z}\dfrac{\partial }{\partial y}+%
\dfrac{\partial }{\partial z}\cdot \alpha _{y} \\ 
\alpha _{x}\dfrac{\partial }{\partial z}+\dfrac{\partial }{\partial x}\cdot
\alpha _{z} & \alpha _{y}\dfrac{\partial }{\partial z}+\dfrac{\partial }{%
\partial y}\cdot \alpha _{z} & \alpha _{z}\dfrac{\partial }{\partial z}+%
\dfrac{\partial }{\partial z}\cdot \alpha _{z}%
\end{pmatrix}%
,  \label{conLPdens}
\end{equation}%
where $\partial /\partial x\cdot \alpha _{y}=\alpha _{y}\partial /\partial
x+\partial \alpha _{y}/\partial x$. Assuming $\delta \mathfrak{K/}\delta
\alpha =X_{K},$ the Lie-Poisson equations on $\mathfrak{X}_{con}^{\ast
}\left( \mathcal{M}\right) $ are 
\begin{equation}
\dot{\alpha}=J_{LP}\left( \alpha \right) X_{K}=-ad_{X_{K}}^{\ast }\alpha =-%
\mathcal{L}_{X_{K}}\alpha -\left( div_{d\mu }X_{K}\right) \alpha .
\label{conalp}
\end{equation}
\end{proposition}

The Lie-Poisson bracket on the dual space $Den\left( \mathcal{M}\right) $ of 
$\mathcal{F}\left( \mathcal{M}\right) $, as defined by Eq.(\ref{condenmom}),
is 
\begin{equation}
\left\{ \mathfrak{H},\mathfrak{K}\right\} \left( L\right) =\int_{\mathcal{M}%
}L\left\{ \frac{\delta \mathfrak{H}}{\delta L},\frac{\delta \mathfrak{K}}{%
\delta L}\right\} _{c}d\mu =\int_{\mathcal{M}}L\left\{ H,K\right\} _{c}d\mu ,
\label{conLP2}
\end{equation}%
where $\delta \mathfrak{H/}\delta L=H,$ $\delta \mathfrak{K}/\delta L=K\in 
\mathcal{F}\left( \mathcal{M}\right) $ and $d\mu =d\sigma \wedge \sigma $.

\begin{proposition}
The Hamiltonian operator $J_{LP}\left( L\right) $ for the Lie Poisson
bracket in Eq.(\ref{conLP}) is 
\begin{equation}
J_{LP}\left( L\right) =X_{L}+\left( 4L+\frac{\partial L}{\partial z}\right) 
\frac{\partial }{\partial z},  \label{conLPdens2}
\end{equation}%
and the Lie-Poisson equation on $Den\left( \mathcal{M}\right) $ becomes 
\begin{equation}
\dot{L}=-\left\{ L,K\right\} _{c}-2div_{d\mu }\left( X_{K}\right) L.
\label{coneq}
\end{equation}

\begin{proof}
The verification of the Hamiltonian operator in Eq.(\ref{conLPdens2}) is a
straightforward calculation which follows directly from the definition of
the Lie-Poisson bracket in Eq.(\ref{conLP}). To obtain the Lie-Poisson
equation we compute the coadjoint action negative of which is the required
equation. By definition%
\begin{eqnarray}
\left\langle ad_{K}^{\ast }L,H\right\rangle  &=&\left\langle
L,ad_{K}H\right\rangle =\left\langle L,\left\{ K,H\right\} _{c}\right\rangle 
\notag \\
&=&-\int_{\mathcal{N}}L\left\{ H,K\right\} _{c}d\mu =-\int_{\mathcal{N}%
}L\left( X_{K}\left( H\right) +\frac{\partial K}{\partial z}H\right) d\mu  
\notag \\
&=&\int_{\mathcal{N}}\left( X_{K}\left( L\right) +div_{d\mu }\left(
X_{K}\right) L-\frac{\partial K}{\partial z}L\right) Hd\mu   \notag \\
&=&\int_{\mathcal{N}}\left( \left\{ L,K\right\} _{c}-\frac{\partial K}{%
\partial z}L+div_{d\mu }\left( X_{K}\right) L-\frac{\partial K}{\partial z}%
\right) Hd\mu   \notag \\
&=&\int_{\mathcal{N}}\left( \left\{ L,K\right\} _{c}+2div_{d\mu }\left(
X_{K}\right) L\right) Hd\mu ,
\end{eqnarray}%
where we use integration by parts at the third step and the identities 
\begin{equation}
\left\{ H,K\right\} _{c}=X_{K}\left( H\right) +\frac{\partial K}{\partial z}%
H=-X_{H}\left( K\right) -\frac{\partial H}{\partial z}K
\end{equation}%
at the second and fourth steps. 
\end{proof}
\end{proposition}

The equation of motion $\dot{L}=-ad_{K}^{\ast }L$ is the kinetic equation of
contact particles in density formulation.

\begin{proposition}
The Hamiltonian differential operators $J_{LP}\left( \alpha \right) $ in Eq.(%
\ref{conLPdens}) and $J_{LP}\left( L\right) $ in Eq.(\ref{conLPdens2}) are
related by%
\begin{equation}
HJ_{LP}\left( L\right) K=-X_{H}J_{LP}\left( \alpha \right) X_{K}\text{ \ \ \
(mod }div\text{).}  \label{conLPeq}
\end{equation}
\end{proposition}

We now obtain dynamics of contact particles by the methods of previous
sections. Let $\left( \mathcal{M},\sigma \right) $ be a contact manifold and
consider the contact vector field $X_{K}$ in Eq.(\ref{convec}). Its complete
cotangent lift is 
\begin{equation}
X_{K}^{c\ast }=X_{K}+\left( \Upsilon \frac{\partial K}{\partial z}+\Psi 
\frac{\partial K}{\partial x}\right) \frac{\partial }{\partial \alpha _{x}}%
+\left( \Phi +\Psi \right) \left( \frac{\partial K}{\partial y}\frac{%
\partial }{\partial \alpha _{y}}+\frac{\partial K}{\partial z}\frac{\partial 
}{\partial \alpha _{z}}\right) 
\end{equation}%
where we use the following abbreviations%
\begin{equation}
\Upsilon =\alpha _{x}\left( 1+x\frac{\partial }{\partial x}\right) ,\text{ \
\ }\Psi =\alpha _{y}\frac{\partial }{\partial x}-\alpha _{x}\frac{\partial }{%
\partial y}-x\alpha _{z}\frac{\partial }{\partial x},\text{ \ \ }\Phi
=x\alpha _{x}\frac{\partial }{\partial z}+\alpha _{z}
\end{equation}%
and the induced coordinates $\left( x,y,z,\alpha _{x},\alpha _{y},\alpha
_{z}\right) $ on $T^{\ast }\mathcal{N}$. $X_{K}^{c\ast }$ is a canonically
Hamiltonian vector field. The vertical representative $VX_{K}^{c\ast }$ of $%
X_{K}^{c\ast }$ is 
\begin{eqnarray}
VX_{K}^{c\ast } &=&\left( \Upsilon \frac{\partial K}{\partial z}+\Psi \frac{%
\partial K}{\partial x}-X_{K}\left( \alpha _{x}\right) \right) \frac{%
\partial }{\partial \alpha _{x}}+\left( \left( \Phi +\Psi \right) \frac{%
\partial K}{\partial y}-X_{K}\left( \alpha _{y}\right) \right) \frac{%
\partial }{\partial \alpha _{y}}  \notag \\
&&+\left( \left( \Phi +\Psi \right) \frac{\partial K}{\partial z}%
-X_{K}\left( \alpha _{z}\right) \right) \frac{\partial }{\partial \alpha _{z}%
},
\end{eqnarray}%
with $X_{K}\left( \alpha _{x}\right) $ denoting the action of $X_{K}$ on $%
\alpha _{x}$. To obtain the equations of motion for the momentum variables,
one needs to add the divergence term, that is,%
\begin{equation}
\dot{\alpha}=VX_{K}^{c\ast }\left( \alpha \right) -\left( div_{d\mu
}X_{K}\right) \alpha .  \label{conalp2}
\end{equation}%
It can be checked that Eq.(\ref{conalp2}) and Eq.(\ref{conalp}) are equal.
In coordinates, the system of equations in Eq.(\ref{conalp2}) takes the form 
\begin{eqnarray}
\dot{\alpha}_{x} &=&\Upsilon \frac{\partial K}{\partial z}+\Psi \frac{%
\partial K}{\partial x}-X_{K}\left( \alpha _{x}\right) +2\frac{\partial K}{%
\partial z}\alpha _{x}  \notag \\
\dot{\alpha}_{y} &=&\left( \Phi +\Psi \right) \frac{\partial K}{\partial y}%
-X_{K}\left( \alpha _{y}\right) +2\frac{\partial K}{\partial z}\alpha _{y} 
\notag \\
\dot{\alpha}_{z} &=&\left( \Phi +\Psi \right) \frac{\partial K}{\partial z}%
-X_{K}\left( \alpha _{z}\right) +2\frac{\partial K}{\partial z}\alpha _{z}.
\label{connc}
\end{eqnarray}%
Substituting $L$ in Eq.(\ref{el}) to the system of Eqs.(\ref{connc}) we
obtain the evolution of the density of contact particles as given by Eq.(\ref%
{coneq}).

\end{document}